\documentclass[a4paper,journal]{IEEEtran} % use the `journal` option for ITherm conference style
\IEEEoverridecommandlockouts
\usepackage[left=0.625in,right=0.625in,top=0.7in,bottom=1.1in]{geometry}
\usepackage[dvipsnames]{xcolor}
\usepackage{cite}
\usepackage{amsmath,amssymb,amsfonts}
\usepackage{graphicx}
\usepackage{textcomp}
\usepackage{xcolor}
\usepackage{float}
\usepackage{cite}
\usepackage{placeins}
\usepackage{textcomp}
\usepackage{multirow}
\usepackage{xcolor}
\usepackage{wrapfig}
\usepackage{algpseudocode}
\usepackage{algorithm}
\usepackage{subcaption}
\usepackage{comment}

\usepackage[noabbrev]{cleveref}

\usepackage{tabularx, multirow,multicol,colortbl,array}
\usepackage{graphicx,color, array}
\usepackage[dvipsnames]{xcolor}
\usepackage{float}
\usepackage{comment,psfrag}
\usepackage{enumitem}
\usepackage{tikz,tikzscale,bigints,pgfplots}
\pgfplotsset{compat=1.18}
\usetikzlibrary{positioning, arrows.meta, colorbrewer, calc, fit, matrix}
\usetikzlibrary{intersections}
\usepgfplotslibrary{fillbetween}
\usepackage{multirow}

\usepackage{import}
\usepackage{pgfplots}
\usetikzlibrary{calc}
\usetikzlibrary{spy}
\usetikzlibrary{matrix}
\usetikzlibrary{backgrounds}
\usetikzlibrary{shapes.misc, positioning}
\usetikzlibrary{shadows,arrows,shapes,calc,positioning,decorations.pathreplacing,fit}
\pgfplotsset{compat = 1.18}

\def\BibTeX{{\rm B\kern-.05em{\sc i\kern-.025em b}\kern-.08em
		T\kern-.1667em\lower.7ex\hbox{E}\kern-.125emX}}

\makeatletter
\def\endthebibliography{%
  \def\@noitemerr{\@latex@warning{Empty `thebibliography' environment}}%
  \endlist
}
\makeatother

\begin{document}

\pgfplotsset{
    standard/.style={
    axis line style = thick,
    grid = both,
    }
}

%% Remove page number
\pagestyle{empty}

\setlength{\abovecaptionskip}{1pt}
\setlength{\belowcaptionskip}{1pt}
% \captionsetup[subfigure]{skip=3pt}

\title{Robust Blind Channel Estimation for Bursty Impulsive Noise with a Constrained EM Approach

}

\author{%%%% author names
    \IEEEauthorblockN{ Chin-Hung Chen$^{\star}$}% first author
    , \IEEEauthorblockN{Ivana Nikoloska$^{\star}$}%
    , \IEEEauthorblockN{Wim van Houtum$^{\star\dagger}$}% 
    , \IEEEauthorblockN{Yan Wu$^{\dagger}$}% 
    , \IEEEauthorblockN{Boris Karanov$^{\ddagger}$}% 
    , and \IEEEauthorblockN{Alex Alvarado$^{\star}$}% 
    \\%%%% author affiliations
    \IEEEauthorblockA{\textit{$^{\star}$Information and Communication Theory Lab, Eindhoven University of Technology, The Netherlands}}\\
    \IEEEauthorblockA{\textit{$^{\dagger}$NXP Semiconductors, Eindhoven, The Netherlands}}\\
    \IEEEauthorblockA{\textit{$^{\ddagger}$Communications Engineering Lab, Karlsruhe Institute of Technology, Germany}}\\
    %%%% corresponding author contact details
    \IEEEauthorblockA{c.h.chen@tue.nl}
}

\maketitle
\thispagestyle{empty}

\begin{abstract}
    Impulsive noise (IN) commonly generated by power devices can severely degrade the performance of high-sensitivity wireless receivers. Accurate channel state information (CSI) knowledge is essential for designing optimal maximum a posteriori detectors. This paper examines blind channel estimation methods based on the expectation-maximization (EM) algorithm tailored for scenarios impacted by bursty IN, which can be described by the Markov-Middleton model. We propose a constrained EM algorithm that exploits the trellis structure of the IN model and the transmitted binary phase-shift keying (BPSK) symbols. By enforcing shared variance among specific trellis states and symmetry in the transition matrix, the proposed constrained EM algorithm adapted for the bursty IN channel has an almost two times faster convergence rate and better estimation performance than the standard EM approach. We comprehensively evaluate the robustness of both standard and constrained EM estimators under different types of CSI uncertainties. The results indicate that the final estimations of both EM estimators are robust enough to mismatch Markov-Middleton model parameters. However, as the level of CSI uncertainty increases, the convergence rate decreases.

\end{abstract}


%% Main Texts------------------------------------------------
\section{Introduction} \label{sec:intro}
This paper examines blind estimation techniques for channels affected by bursty impulsive noise (IN) generated by power devices, such as power lines and DC-DC converters. It is well established that these electronic components, found in power substations and electric vehicles (e.g., \cite{Sacuto14, CHC24_2}), emit consecutive high-energy spikes, which can seriously degrade the performance of high-sensitivity wireless communication receivers \cite{Wim22, Landa15, Shan11}. Therefore, several studies have concentrated on modeling bursty IN to develop robust receivers equipped with specialized detectors. The well-known bursty IN models including the two-state Markov-Gaussian model \cite{Dario09} and the finite-state Markov-Middleton model \cite{MMA} adopted the hidden-Markov model (HMM) to address the correlation between IN samples. Our previous study \cite{CHC24_2} also discovered that a Markov-Middleton model effectively represents the EV-induced interference observed in field tests.

Conventional detectors for handling IN, such as clipping \cite{Ndo10} and blanking \cite{Oh17}, are widely adopted due to their simple implementation and low computational complexity. However, their performance is inherently suboptimal. In contrast, the design of optimal IN detectors, which utilize the statistical characteristics of the underlying channel and employ maximum a posteriori (MAP) detection, has been extensively investigated across various IN models and transmission formats \cite{Dario09, MMA, Mitra10, CHC24_3}. In \cite{Alam20}, a thorough comparison of various IN detectors is presented. Optimal MAP detectors can be implemented using the Bahl-Cocke-Jelinek-Raviv (BCJR) algorithm \cite{bcjr74} to compute posterior probabilities; however, these detectors are highly sensitive to channel state information (CSI) \cite{Dario09, MMA}. Therefore, accurate parameter estimation is crucial for the optimal IN detectors. 
        
Channel estimators that use the expectation maximization (EM) algorithm have received considerable attention in the field of wireless communications, largely due to their ability to iteratively improve parameter estimates without depending on training data. This advantage has made the EM algorithm a powerful method for blind channel estimation in dynamic and noisy environments. In recent decades, the EM estimator has become a mature topic for wireless channels affected by intersymbol interference (ISI) \cite{Ghosh92, Kaleh94, Lopes01_2, Schmid24, CHC25}. Although similar EM-based methods have been adapted for IN channels, most studies focus on memoryless IN, where noise samples are independent, and the state transition probability is considered trivial \cite{Zabin91, Chen19, Awino19}. In \cite{Vannucci19} and \cite{Mirbadin21}, an approximate message-passing algorithm was proposed to estimate the correlated transmitted symbols and the bursty IN in parallel through a loopy factor graph. Recently, \cite{karanov24} introduced an EM-based estimator tailored for bursty IN channels with uncorrelated transmitted symbols. This implementation combines transmitted symbols and IN into one loop-free factor graph, facilitating the estimation of channel likelihood and state transition probabilities.

Our study builds upon the work of \cite{karanov24} by proposing a constrained EM algorithm that enforces shared variance among specific trellis states and ensures symmetry in the transition probabilities. Compared to the standard EM algorithm adopted in \cite{karanov24}, we demonstrate that the constrained EM estimator for the bursty IN channels is more robust and converges faster. Moreover, we perform a thorough robustness analysis to evaluate the estimation performance of both standard and constrained EM estimators across various initialization errors, reflecting the different levels of CSI uncertainties that can occur in practice.

This paper is structured as follows: Sec.~\ref{sec:sys_model} summarizes the transmission over the Markov-Middleton model. Sec.~\ref{sec:hmm} illustrates the standard and constrained EM algorithms. In Sec.~\ref{sec:results}, we present the numerical simulation for the considered EM estimators. Finally, Sec.~\ref{sec:conc} summarizes the paper.

\section{System Model} \label{sec:sys_model}

%% Tramsmission configuration ===================
In this paper, we assume a binary phase-shift keying (BPSK) modulation, where a symbol mapper maps the binary bits $u_t \in \{0, 1\}$ to a BPSK symbol $x_t \in \{-1,+1\}$. The symbol time step is denoted by $t$. For transmission over a Markov-Middleton IN channel, the received observation sequence $\mathbf{y}_1^T$ of length $T$ can be described by $\mathbf{y}_1^T = \mathbf{x}_1^T + \mathbf{n}_1^T$,
where we use boldface letters to denote a sequence, with subscripts and superscripts denoting the start and end of the sequence. The statistical properties of noise samples $n_t$ are completely determined by the noise state realizations at time $t$, which is defined as $w_t \in \mathcal{W}=\{0,1,\cdots, W-1\}$, where $w_t=0$ represents the background nose state, and $W$ is the total number of noise states. The generic probability density function (PDF) of the Markov-Middleton noise sample $n_t$ can be written as the finite Gaussian mixture model \cite{MMA} as
\begin{align}
    p(n_t) &= \sum_{j\in \mathcal{W}} P(w_t=j)\cdot p(n_t|w_t=j) \nonumber\\
    &=\sum_{j\in \mathcal{W}}\frac{P(w_t=j)}{\sqrt{2\pi \sigma^2_{w_j}}}\exp{\left(-\frac{n_t^2}{2\sigma^2_{w_j}}\right)}, \label{eq:MMApdf2}
\end{align}
where we use $p(\cdot)$ to denote a PDF for continuous random variables and $P(\cdot)$ to denote a probability mass function for discrete random variables. In \eqref{eq:MMApdf2}, the probability of being in state $j$ is written as
\begin{align}\label{eq:pm2}
    P(w_t=j) = \frac{e^{-A}A^j/j!}{\sum_{i\in\mathcal{W}} e^{-A}A^i/i!},
\end{align}
where $A$ indicates the impulsive index; as $A$ increases, the amount of noise in the IN states ($j>0$) increases. The noise variance $\sigma^2_{w_j}$ at a specific state $w_t=j$ is defined as
\begin{align}
    \sigma^2_{w_j} = \left( 1 + \frac{j\Lambda}{A} \right) \sigma^2_{w_0}, \nonumber
\end{align}
where $\sigma^2_{w_0}$ is the background noise variance and $\Lambda = \sigma_I^2/\sigma_B^2$ is the impulsive-to-background average noise power ratio.\footnote{In the original paper \cite{MMA}, the power ratio is defined as $\Gamma=\sigma_B^2/\sigma_I^2$.} 

\begin{figure}[t]\label{fig:MMA_noise}
\centering
    \subfloat[$A=0.1, \Lambda = 1, r=0$\label{fig:MMA_noise1}]{%
        \centering
        {	
\begin{tikzpicture}
    \begin{axis}[ 
        axis y line*=right,
        axis line style = thick,
        axis x line=none,
		grid = both,
		name = p1,
		% table/col sep=comma,
		width=\columnwidth,
		height=.33\columnwidth, 
  		xmin = 1, xmax=500,
	   	ymin = 0.9, ymax=2,
  		ytick = {1,2}, yticklabels={0,1},
            ylabel = {$w_t$},
            ylabel style={color=orange},
            yticklabel style={color=orange}, 
		font=\footnotesize,
        ]
		\addplot[ycomb, very thin, color=orange!50, mark=*, mark size=1, mark options={fill=white, semithick}] 
		table[x expr=\thisrowno{0}, y expr=\thisrowno{2}] 
		{data/noise_A01L1r0.txt};
	\end{axis}
 
	\begin{axis}[ 
        axis line style = thick,
		name = p1,
		% table/col sep=comma,
		width=\columnwidth,
		height=.33\columnwidth, 
  		xmin = 1, xmax=500,
	   	ymin = 0, ymax=12,
            tick style={draw=none},
  		ytick = {0,4,...,12}, yticklabels={0,4,...,12},
            xtick = {0, 100,200,300,...,1000}, xticklabels={0,100,...,1000},
            xlabel = Sample, ylabel = {$|n_t|$},
		font=\footnotesize,
 ]
		\addplot[color=black, semithick] 
		table[x expr=\thisrowno{0}, y expr=\thisrowno{1}] 
		{data/noise_A01L1r0.txt};
		
	\end{axis}

\end{tikzpicture}    } 
    }
    \hfill
    \subfloat[$A=0.3, \Lambda = 10, r=0.9$\label{fig:MMA_noise2}]{%
        \centering
       {	
\begin{tikzpicture}
 
	\begin{axis}[ 
        axis y line*=right,
        axis line style = thick,
        axis x line=none,
		grid = both,
		name = p1,
		% table/col sep=comma,
		width=\columnwidth,
		height=.33\columnwidth, 
  		xmin = 1, xmax=500,
	   	ymin = .9, ymax=2,
  		ytick = {1,2}, yticklabels={0,1},
            ylabel = {$w_t$},
            ylabel style={color=orange},
            yticklabel style={color=orange}, 
		font=\footnotesize,
        ]
		\addplot[ycomb, very thin, color=orange!50, mark=*, mark size=1, mark options={fill=white, semithick}] 
		table[x expr=\thisrowno{0}, y expr=\thisrowno{2}] 
		{data/noise_A03L10r9.txt};
		
	\end{axis}

	\begin{axis}[ 
            axis line style = thick,
		name = p1,
		% table/col sep=comma,
		width=\columnwidth,
		height=.33\columnwidth, 
  		xmin = 1, xmax=500,
	   	ymin = 0, ymax=12,
            tick style={draw=none},
  		ytick = {0,4,...,12}, yticklabels={0,4,...,12},
            xtick = {0,100,...,1000}, xticklabels={0,100,...,1000},
            xlabel = Sample, ylabel = {$|n_t|$},
		font=\footnotesize,
        ]
		\addplot[color=black, semithick] 
		table[x expr=\thisrowno{0}, y expr=\thisrowno{1}] 
		{data/noise_A03L10r9.txt};
		
	\end{axis}
 
\end{tikzpicture}    } 
    }
    \caption{Noise realization from a $2$-state Markov-Middleton model ($W=2$) with $\sigma^2_{w_0}=1$ for $A=0.1$ and $A=0.3$, $\Lambda=1$ and $\Lambda=10$, and $r=0$ and $ r=0.9$. The orange vertical bar denotes the index $j$ of the state realizations \( w_t = j \) for generating the corresponding noise samples.}
\end{figure}
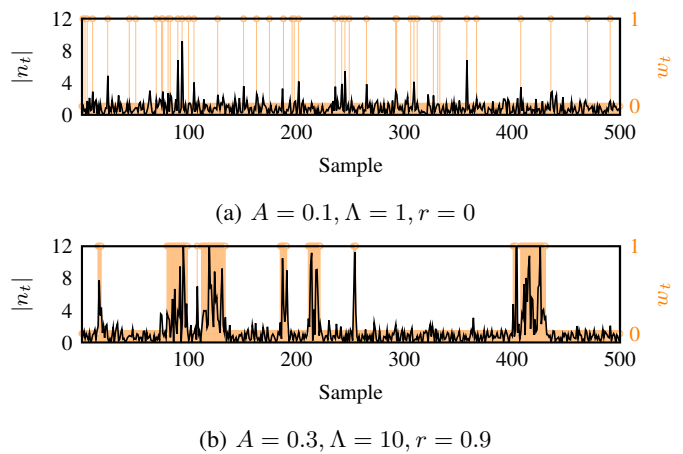

To describe the burstiness (memory) of the noise observed in practice, \cite{MMA} introduces a parameter $r\in[0, 1)$ to establish a correlation between consecutive noise samples. Specifically, the time-invariant state transition matrix of the Markov-Middleton noise model is constructed as 
\begin{align}\label{eq:P}
    P_{w_{ij}} &= P(w_{t}=j | w_{t-1} = i) \nonumber \\ 
    &= \begin{cases}
        r + (1-r) \cdot P(w_t=j), \quad i=j\\
        (1-r) \cdot P(w_t=j),  \quad \text{otherwise}.
    \end{cases}
\end{align}
From \eqref{eq:P}, we can see that the larger the correlation parameter $r$, the more likely the following noise sample will stay in the same state. That is, the burst behavior is more prominent. In Figs.~\ref{fig:MMA_noise1} and \ref{fig:MMA_noise2}, we show two-state ($W=2$) Markov-Middleton noise realizations for two different parameter sets. Fig.~\ref{fig:MMA_noise1} has an impulsive state probability $P(w_t=1)\approx 0.09$, whereas Fig.~\ref{fig:MMA_noise2} is featured with $P(w_t=1)\approx 0.23$ as derived from $A=0.1$ and $A=0.3$, respectively. Compared to Fig. \ref{fig:MMA_noise1}, Fig. \ref{fig:MMA_noise2} exhibits a higher level of impulsive noise power ($\Lambda=10$) and a stronger noise correlation ($r=0.9$), where the noise tends to remain in the same state for consecutive samples. In contrast, Fig. \ref{fig:MMA_noise1} shows independent noise samples with a correlation of $r=0$. The noise states in this figure are randomly selected from the set $w_t \in \mathcal{W}$, following the probability defined in \eqref{eq:pm2}.

\section{EM-based estimators design}\label{sec:hmm}

In this section, we provide an overview of the EM algorithm for finite-state channels with memory. This leads to the well-known Baum-Welch algorithm \cite{Baum_72}, which is a specialization of the EM algorithm for HMMs. More specifically, we utilize the EM algorithm for data that can be described by a Gaussian mixture model, incorporating an additional Markov process to account for state transitions. The objective of the EM algorithm is to iteratively identify the maximum likelihood solutions based on the distributions of both observed and latent variables so as to maximize the model evidence, which can be written as
\begin{align} \label{eq:evidence}
    p(\mathbf{y}_1^T | \theta) = \sum_{\mathbf{s}_1^T} p(\mathbf{y}_1^T, \mathbf{s}_1^T | \theta).
\end{align}
Here, $s_t$ denotes the latent variable (state) realizations and $\theta$ represents the set of model parameters, which will be discussed later. We can interpret \eqref{eq:evidence} as the likelihood of the observation sequence given the set of model parameters $\theta$.

In the following, we will first introduce the HMM representation for BPSK transmission over the Markov-Middleton IN channel. Next, we will present the E-step of the EM algorithm, which focuses on determining the posterior distribution of the latent variable given the observation sequence. This step can be accomplished using the BCJR algorithm. Following this, the M-step consists of maximum likelihood parameter estimations based on the posterior probabilities computed in the E-step.

\subsection{HMM structure for BPSK over bursty IN channel} \label{sec: hmm}
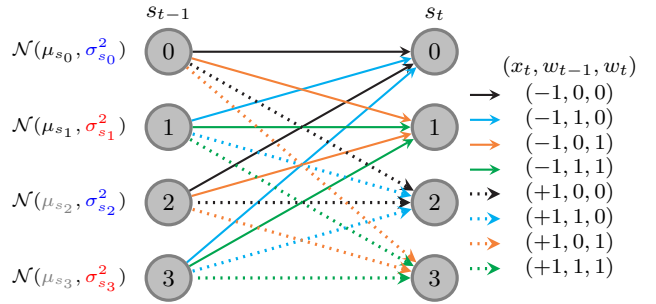
\begin{figure}[t]
    \centering
    {\tikzstyle{state}=[shape=circle,draw=black!50, fill=gray!50, font=\small, very thick]
\tikzstyle{edge}=[stealth-, thick]

\tikzstyle{edge1}=[edge, Black]
\tikzstyle{edge2}=[edge, Cyan]
\tikzstyle{edge3}=[edge, Orange]
\tikzstyle{edge4}=[edge, Green]
\tikzstyle{edge5}=[edge, Black, dotted,very thick]
\tikzstyle{edge6}=[edge, Cyan, dotted,very thick]
\tikzstyle{edge7}=[edge, Orange, dotted,very thick]
\tikzstyle{edge8}=[edge, Green, dotted,very thick]
\def\ttsize{\scriptsize}

\begin{tikzpicture}[]

\node  (s')         at (0,5.5)  {\small$s_{t-1}$};
\node  (s1)         at (-1.3,5) {\ttsize$\mathcal{N}(\textcolor{black}{\mu_{s_0}},\textcolor{blue}{\sigma^2_{s_0}})$};
\node               at (-1.3,4) {\ttsize$\mathcal{N}(\textcolor{black}{\mu_{s_1}},\textcolor{red}{\sigma^2_{s_1}})$};
\node               at (-1.3,3) {\ttsize$\mathcal{N}(\textcolor{gray}{\mu_{s_2}},\textcolor{blue}{\sigma^2_{s_2}})$};
\node               at (-1.3,2) {\ttsize$\mathcal{N}(\textcolor{gray}{\mu_{s_3}},\textcolor{red}{\sigma^2_{s_3}})$};

% 1st column-----------------------------------
\node[state] (s1_1) at (0,5) {$0$};
\node[state] (s2_1) at (0,4) {$1$};
\node[state] (s3_1) at (0,3) {$2$};
\node[state] (s4_1) at (0,2) {$3$};

% 2nd column----------------------------------
\node               at (3.5,5.5) {\small$s_t$};
\node[state] (s1_2) at (3.5,5) {$0$}
    edge[edge1] (s1_1)%node[above] {\scriptsize$p(y_t | s_t) \cdot p(s_t|s_{t-1})$} (s1_1) 
    edge[edge2] (s2_1)
    edge[edge1] (s3_1) 
    edge[edge2] (s4_1);
\node[state] (s2_2) at (3.5,4) {$1$}
    edge[edge3] (s1_1)%node[above] {\scriptsize$p(y_t | s_t) \cdot p(s_t|s_{t-1})$} (s1_1) 
    edge[edge4] (s2_1)
    edge[edge3] (s3_1) 
    edge[edge4] (s4_1);

\node[state] (s3_2) at (3.5,3) {$2$}
    edge[edge5] (s1_1)%node[above] {\scriptsize$p(y_t | s_t) \cdot p(s_t|s_{t-1})$} (s1_1) 
    edge[edge6] (s2_1)
    edge[edge5] (s3_1) 
    edge[edge6] (s4_1);
    
\node[state] (s4_2) at (3.5,2) {$3$}
    edge[edge7] (s1_1)%node[above] {\scriptsize$p(y_t | s_t) \cdot p(s_t|s_{t-1})$} (s1_1) 
    edge[edge8] (s2_1)
    edge[edge7] (s3_1) 
    edge[edge8] (s4_1);

\node[
    anchor=west,
    fill=white,
    font=\footnotesize,
    yshift=-1.5em
    ] at (s2_2.east) {
    \begin{tabular}{@{} c@{}c@{}}
        % \multicolumn{2}{c}{Trellis Input} \\ 
        & $(x_t, w_{t-1}, w_t)$ \\ 
        \begin{tikzpicture}
            \draw[edge1,-stealth] (.4,0) -- (.8,0);
        \end{tikzpicture} & \raisebox{-0.3ex}{$(-1, 0, 0)$} \\
        \begin{tikzpicture}
            \draw[edge2,-stealth] (.4,0) -- (.8,0);
        \end{tikzpicture} & \raisebox{-0.3ex}{$(-1, 1, 0)$} \\
                \begin{tikzpicture}
            \draw[edge3,-stealth] (.4,0) -- (.8,0);
        \end{tikzpicture} & \raisebox{-0.3ex}{$(-1, 0, 1)$} \\
                \begin{tikzpicture}
            \draw[edge4,-stealth] (.4,0) -- (.8,0);
        \end{tikzpicture} & \raisebox{-0.3ex}{$(-1, 1, 1)$} \\
                \begin{tikzpicture}
            \draw[edge5,-stealth] (.4,0) -- (.8,0);
        \end{tikzpicture} & \raisebox{-0.3ex}{$(+1, 0, 0)$} \\
                \begin{tikzpicture}
            \draw[edge6,-stealth] (.4,0) -- (.8,0);
        \end{tikzpicture} & \raisebox{-0.3ex}{$(+1, 1, 0)$} \\
                \begin{tikzpicture}
            \draw[edge7,-stealth] (.4,0) -- (.8,0);
        \end{tikzpicture} & \raisebox{-0.3ex}{$(+1, 0, 1)$} \\
                \begin{tikzpicture}
            \draw[edge8,-stealth] (.4,0) -- (.8,0);
        \end{tikzpicture} & \raisebox{-0.3ex}{$(+1, 1, 1)$} \\
    \end{tabular}
    };
\end{tikzpicture}} 
    \caption{Trellis section of a BPSK transmission over a 2-state ($W=2$) IN model, where each state $s_t=j$ denote a single Gaussian component $\mathcal{N}(\mu_{s_j}, \sigma^2_{s_j})$. The state transitions are driven by ($x_t$, $w_{t-1}$, $w_{t}$) with probabilities $P_{s_{ij}}$. We use the same color to indicate metrics that share identical values.} 
    \label{fig:trellis_sec}
\end{figure}

An HMM is a statistical model that captures the temporal behavior of a sequence of states $\mathbf{s}_1^T$ from a Markov process via a sequence of state-dependent observations $\mathbf{y}_1^T$. We assume the state realization $s_t$ is drawn from a time-invariant finite-state space $\mathcal{S}=\{0,1,\cdots,S-1\}$. Similar to \cite{CHC24_3} and \cite{karanov24}, we combine the transmitted BPSK symbol and IN state into a single HMM model. In this model, the total number of states is $S=2 \times W$, where each state realization $s_t=j$ serves as an indicator to identify which Gaussian component $\mathcal{N}(\mu_{s_j}, \sigma^2_{s_j})$ generates the corresponding observed data $y_t$. We can therefore write the likelihood function as
\begin{align}\label{eq:lik_ys}
    p(y_t|s_t = j) &= \frac{1}{\sqrt{2\pi \sigma^2_{s_j}}}\exp{\left(-\frac{(y_t - \mu_{s_j})^2}{2\sigma^2_{s_j}}\right)}, 
\end{align}
where $\mu_{s_j} \in \{-1,+1\}$ is dictated by the BPSK symbol energy and $\sigma^2_{s_j} \in \{\sigma^2_{w_0},\sigma^2_{w_1},\cdots,\sigma^2_{w_{W-1}}\}$ is determined by the IN state. The state-transition probability $ P_{s_{ij}} $ of moving from state $i$ to state $j$ is determined by both the Markov-Middleton noise transition matrix \eqref{eq:P} and the prior symbol probability $P(x)$ and can therefore be expressed as 
\begin{align}\label{eq:ptran_s}
    P_{s_{ij}} &= P(s_t = j | s_{t-1} = i ) \nonumber \\
    &= P_{w_{uv}} \cdot P(x), \quad w_t = v, \quad w_{t-1}=u,
\end{align}
where $u,v \in \mathcal{W}$. Note that we assume the transmitted BPSK symbols are independent, where $P(x)=1/2$. In Fig.~\ref{fig:trellis_sec}, we present a trellis section example for BPSK transmission over a 2-state ($W=2$) Markov-Middleton model, where each state represents a single Gaussian distribution $\mathcal{N}(\mu_{s_j}, \sigma^2_{s_j})$, and the connection (edges) shows the state transitions characterized by $P_{s_{ij}}$. 

Given the total number of states $S$, the parameter set that we aim to estimate through the EM algorithm is
\begin{align}\label{eq:hmmpara}
    \theta = \{\mu_{s_j}, \sigma^2_{s_j}, P_{s_{ij}}\}, \quad i,j \in \mathcal{S}.
\end{align}
%%--------------------------------------------------------------
\subsection{E-step: posterior distribution inference}
For a general EM algorithm, the E-step aims at finding a surrogate distribution $q(s_t)$ that maximizes the evidence lower bound (ELBO) with the model parameter $\theta$ held fixed. It can be proved that this maximization is achieved when the surrogate distribution is set equal to the posterior distribution \cite{Bishop}, namely $q(s_t) = P(s_t|\mathbf{y}_1^T, \theta^{(l)})$. Here, we use $\theta^{(l)}$ to denote the parameters set estimated from iteration $l$.

In our application, the Markovian structure of the latent state allows an efficient BCJR algorithm for computing the joint distribution of the states at a specific time instant given the full observation sequence as
\begin{align} 
     p(s_t&,  \mathbf{y}^T_1| \theta^{(l)}) = \sum_{s_{t-1}\in\mathcal{S}}{p(s_{t}, s_{t-1}, \mathbf{y}^T_1 | \theta^{(l)})} \label{eq:map1} \\ 
    & = \sum_{s_{t-1}\in\mathcal{S}}  \underbrace{p(\mathbf{y}_{t+1}^{T} | s_{t})}_{\beta(s_t)} \cdot \underbrace{p(y_t, s_t | s_{t-1})}_{\gamma(s_t,s_{t-1})}  \cdot \underbrace{p(s_{t-1}, \mathbf{y}_1^{t-1})}_{\alpha(s_{t-1})} \label{eq:map2},
\end{align}
where we omit the condition on $\theta^{(l)}$ in \eqref{eq:map2} to simplify the notation. In \eqref{eq:map2}, $\alpha(s_{t-1})$, $\beta(s_t)$, and $\gamma(s_t,s_{t-1})$ are commonly used to denote forward recursion $p(s_{t-1}, \mathbf{y}_1^{t-1})$, backward recursion $p(\mathbf{y}_{t+1}^{T} | s_{t})$, and branch metric $p(y_t, s_t | s_{t-1})$, respectively. The branch metric $\gamma$ can be further expressed as 
\begin{align} \label{eq:gamma}
        \gamma(s_{t}, s_{t-1}) =  p(y_t| s_t) \cdot P_{s_{ij}}, 
\end{align}
which can be seen as the multiplication of observation likelihood~\eqref{eq:lik_ys} and the state transition probability~\eqref{eq:ptran_s}.
The forward recursion representation of $\alpha$ can be derived as
\begin{align} \label{eq:siso_alpha}
    \alpha(s_{t}) &= p(s_{t}, \mathbf{y}_1^{t}) = \sum_{s_{t-1}\in\mathcal{S}} \gamma(s_{t}, s_{t-1}) \cdot \alpha(s_{t-1}), 
\end{align}
and the backward recursion $\beta$ can be written as
\begin{align} \label{eq:siso_beta}
    \beta(s_{t-1}) &= p(\mathbf{y}_{t}^{T} | s_{t-1}) = \sum_{s_{t}\in\mathcal{S}} \beta(s_t) \cdot \gamma(s_{t}, s_{t-1}). 
\end{align}
% \begin{align} \label{eq:siso_alpha}
%     \alpha(s_{t}) &= p(s_{t}, \mathbf{y}_1^{t}) = \sum_{s_{t-1}\in\mathcal{S}}p(s_{t},s_{t-1}, y_t, \mathbf{y}_1^{t-1}) \nonumber \\
%     % &= \sum_{s_{t-1}\in\mathcal{S}}p(s_{t},s_{t-1}, y_t, \mathbf{y}_1^{t-1}) \label{eq:siso_alpha2} \\ 
%     &= \sum_{s_{t-1}\in\mathcal{S}} \gamma(s_{t}, s_{t-1}) \cdot \alpha(s_{t-1}),
% \end{align}
% and the backward recursion $\beta$ can be written as
% \begin{align} \label{eq:siso_beta}
%     \beta(s_{t-1}) &= p(\mathbf{y}_{t}^{T} | s_{t-1}) =  \sum_{s_{t}\in\mathcal{S}}p(s_t, \mathbf{y}_{t+1}^{T}, y_t | s_{t-1}) \nonumber \\
%     &= \sum_{s_{t}\in\mathcal{S}} \beta(s_t) \cdot \gamma(s_{t}, s_{t-1}).
% \end{align}
The complete derivation of the BCJR and forward-backward algorithms can be found in the existing literature \cite{HMM} and~\cite{bcjr74}. After obtaining the joint probability \eqref{eq:map1} through \eqref{eq:map2}--\eqref{eq:siso_beta}, we can compute the posterior probability of state $s_t$ given the observation sequence $\mathbf{y}_1^T$ via
\begin{align} \label{eq:pos_st}
    P(s_t | \mathbf{y}^T_1, \theta^{(l)}) = \frac{p(s_t, \mathbf{y}^T_1| \theta^{(l)})}{\sum_{s_t \in \mathcal{S}}p(s_t,  \mathbf{y}^T_1| \theta^{(l)})}.
\end{align}

%---------------------------
%---------------------------

%-----------------------------------
\begin{table*}[t]
    \centering
    \caption{One estimation result for variances and the state transition matrix for BPSK over a 2-state Markov-Middleton channel using standard and constrained EM algorithms. The reference and the initialized parameters are also listed.}    \label{tab:para}
    \renewcommand{\arraystretch}{1.1} % Increase row height
    \setlength{\tabcolsep}{5pt} % Adjust horizontal spacing
    \begin{tabular}{|c|c|c|c|c|c|c|c|c|c|c|c|c|c|c|c|c|}
        \hline
         & \multicolumn{4}{c|}{\shortstack{\textbf{Reference} \\ $(A, \Lambda,r) = (0.3, 10, 0.9)$}} & \multicolumn{4}{c|}{\shortstack{\textbf{Initialization} \\ $(A, \Lambda, r) = (0.1, 1, 0)$}} & \multicolumn{4}{c|}{\shortstack{\textbf{Standard EM} \\ 36 Iterations}} & \multicolumn{4}{c|}{\shortstack{\textbf{Constrained EM} \\ 11 Iterations}} \\ \cline{2-17}
         \hline
        {Variance}& \textcolor{blue}{1} &  \textcolor{red}{34.3} &  \textcolor{blue}{1} &  \textcolor{red}{34.3} & 1 & 11 & 1 & 11 & 0.99 & 35.6 & 1.02 & 31.52 &  \textcolor{blue}{1} &  \textcolor{red}{33.7} &  \textcolor{blue}{1} &  \textcolor{red}{33.7} \\ \cline{2-17}
        \hline
        \multirow{4}{*}{\shortstack{Transition\\Matrix}}& 0.488 & \textcolor{Orange}{0.012} & 0.488 & \textcolor{Orange}{0.012} & 0.455  & 0.045 & 0.455 &   0.045 & 0.504 & 0.006 & 0.481 & 0.009 &  0.489 & \textcolor{Orange}{0.011} & 0.489 & \textcolor{Orange}{0.011} \\ \cline{2-17}
        & \textcolor{Cyan}{0.038} & \textcolor{Green}{0.462} & \textcolor{Cyan}{0.038} & \textcolor{Green}{0.462} & 0.455 & 0.045 & 0.455 & 0.045 & 0.067 & 0.407 & 0.051 & 0.475 & \textcolor{Cyan}{0.042} & \textcolor{Green}{0.458} & \textcolor{Cyan}{0.042} & \textcolor{Green}{0.458} \\ \cline{2-17}
        & 0.488 & \textcolor{Orange}{0.012} & 0.488 & \textcolor{Orange}{0.012} & 0.455 & 0.045 & 0.455 & 0.045 & 0.475 & 0.013 & 0.495 & 0.017 & 0.489 & \textcolor{Orange}{0.011} & 0.489 & \textcolor{Orange}{0.011} \\ \cline{2-17}
        & \textcolor{Cyan}{0.038} & \textcolor{Green}{0.462} & \textcolor{Cyan}{0.038} & \textcolor{Green}{0.462} & 0.455 & 0.045 & 0.455 & 0.045 & 0.018 & 0.439 & 0.038 & 0.505 & \textcolor{Cyan}{0.042} & \textcolor{Green}{0.458} & \textcolor{Cyan}{0.042} & \textcolor{Green}{0.458} \\ \hline
    \end{tabular}
\end{table*}
%---------------------------------

%%--------------------------------------------------------------
\subsection{M-step: maximum likelihood parameter estimation}
Based on the posterior belief calculated in the E-step, the M-step conducts the maximum likelihood estimation of the parameter set $\theta$ in \eqref{eq:hmmpara}. Since each state follows a Gaussian distribution, the maximum likelihood estimation \cite{Bishop, Baum_72, HMM} of the means and variance can be updated through

\begin{align}
    \hat{\mu}_{s_j} = \frac{\sum_{t=1}^{T} P(s_t=j | \mathbf{y}_1^T, \theta^{(l)}) \cdot y_t} {\sum_{t=1}^{T} P(s_t =j | \mathbf{y}_1^T, \theta^{(l)})}, \nonumber
\end{align}
and
\begin{align}\label{eq:HMM_sigma}
		\hat{\sigma}^{2}_{s_j} = \frac{\sum_{t=1}^{T} P(s_t=j | \mathbf{y}_1^T, \theta^{(l)}) \cdot (y_t-\hat{\mu}_{s_j})^2} {\sum_{t=1}^{T} P(s_t = j | \mathbf{y}_1^T, \theta^{(l)} )}.
\end{align}

For the state transition probability estimation, we use the joint distribution $p(s_t,s_{t-1}, \mathbf{y}_1^T | \theta^{(l)})$ from the right-hand side of \eqref{eq:map1} and derive the posterior probability of the consecutive two states $s_t$ and $s_{t-1}$ given the whole observation as
\begin{align}
    P(s_t,s_{t-1}| \mathbf{y}_1^T, \theta^{(l)}) = \frac{p(s_t,s_{t-1}, \mathbf{y}_1^T | \theta^{(l)})}{\sum_{s_t \in \mathcal{S}}p(s_t, \mathbf{y}_1^T | \theta^{(l)})}. \nonumber
\end{align}
Subsequently, the state transition probability $P_{s_{ij}}$ is updated through
\begin{align}\label{eq:ptran_new}
    \hat{P}_{s_{ij}} = \frac{\sum_{t=1}^{T} P(s_t=j,s_{t-1}=i | \mathbf{y}_1^T, \theta^{(l)})}{\sum_{t=1}^{T} \sum_{s_t\in\mathcal{S}} P(s_t, s_{t-1}=i | \mathbf{y}_1^T,\theta^{(l)} )}.
\end{align}

The EM algorithm then advances to the next iteration by using the updated model parameters set 
\begin{align}
    \theta^{(l+1)} = \{\hat{\mu}_{s_j}, \hat{\sigma}^{2}_{s_j}, \hat{P}_{s_{ij}}\}, \quad i,j \in \mathcal{S} \nonumber  
\end{align}
to calculate the posterior belief $p(s_t|\mathbf{y}_1^T, \theta^{(l+1)})$ in the next E-step.

%%--------------------------------------------------------------

\subsection{Constrained M-step}
We have introduced the standard EM algorithm to iteratively optimize the mean, variance, and transition probability for any finite state Gaussian mixture distribution, where an HMM can describe the state transition process. In this subsection, we make use of the HMM trellis structure described in Sec.~\ref{sec: hmm} to refine our EM algorithm design. As the joint trellis is a Kronecker product of the BPSK symbol space $\{-1,+1\}$, which determines the mean $\mu_{s_j}$ of the Gaussian distribution, and the Markov-Middleton noise state space $\mathcal{W}$, which dictates the corresponding variance $\sigma^2_{s_j}$, certain states will share the same Gaussian parameters. Using a two-state Markov-Middleton model shown in Fig.~\ref{fig:trellis_sec} for example, the state pair $(0,1)$ and $(2,3)$ have the same means, where $\mu_{s_0}=\mu_{s_1}=-1$ and $\mu_{s_2}=\mu_{s_3}=+1$. While state pair $(0,2)$ and $(1,3)$ share the same variances, where $\sigma^2_{s_0} = \sigma^2_{s_2} = \sigma^2_{w_0}$ and $\sigma^2_{s_1} = \sigma^2_{s_3} = \sigma^2_{w_1}$. We use the same colors in Fig.~\ref{fig:trellis_sec} and Table~\ref{tab:para} to denote the parameters that share the same value.

Instead of updating the means and variances individually, we propose to force shared parameters among specific states. Using the example in Fig.~\ref{fig:trellis_sec}, we write $\hat{\mu}_{c_0}=\hat{\mu}_{c_1}=\frac{1}{2}(\hat{\mu}_{s_0}+\hat{\mu}_{s_1})$ and $\hat{\mu}_{c_2}=\hat{\mu}_{c_3}=\frac{1}{2}(\hat{\mu}_{s_2}+\hat{\mu}_{s_3})$. Similarly, the variances are updated via $\hat{\sigma}^2_{c_0}=\hat{\sigma}^2_{c_2}=\frac{1}{2}(\hat{\sigma}^2_{s_0}+\hat{\sigma}^2_{s_2})$ and $\hat{\sigma}^2_{c_1}=\hat{\sigma}^2_{c_3}=\frac{1}{2}(\hat{\sigma}^2_{s_1}+\hat{\sigma}^2_{s_3})$
The general constrained M-step for updating the Gaussian means and variances can be represented as
\begin{align}
    \hat{\mu}_{c_j} = \frac{1}{|\mathcal{U}_k|} \sum_{i\in \mathcal{U}_k} \hat{\mu}_{s_i}
\end{align}
$\forall j \in \mathcal{U}_k=\{kW,kW+1,\cdots,kW+W-1\}$, where $k=\{0,1\}$ corresponds to the BPSK symbol cardinality, and
\begin{align}
    \hat{\sigma}^{2}_{c_j} = \frac{1}{|\mathcal{V}_k|} \sum_{i\in \mathcal{V}_k}\hat{\sigma}^{2}_{s_i} 
\end{align}
$\forall j \in \mathcal{V}_k=\{k,k+W\}$, where $k=\{0,1,\cdots,W-1\}$ corresponds to the number of noise states. A similar observation applies to the state transition matrix, as illustrated in Fig.~\ref{fig:trellis_sec}, where edges with the same color indicate identical transition probabilities. We can therefore modify \eqref{eq:ptran_new} and derive the general constrained M-step for updating the transition matrix as 
\begin{align}\label{eq:ptran_new_reg}
    \hat{P}_{c{_{ij}}} =  \frac{1}{|\mathcal{P}_{i'j'}|} \sum_{u,v\in \mathcal{P}_{i'j'}} \hat{P}_{s{_{uv}}},
\end{align}
$\forall i,j \in \mathcal{P}_{i'j'} = \{ (i, j) \mid i \in \{i', i' + W\}, j \in \{j', j' + W\} \}$, where $(i', j') \in \{0,1,\cdots,W-1\}^2$.
% $\forall i,j \in \mathcal{P}_k=\{(i,j) | i\in\{k,k+W\}, j\in\{k,k+W\}\}$, where $k=\{0,1,\cdots,|\mathcal{X}|^2-1\}$.

\section{Numerical Simulations}\label{sec:results}

\begin{figure*}[t]
    \centering
    \subfloat[Noise variance\label{fig:mse_var}]{%
        \centering
        {\begin{tikzpicture}
	\begin{semilogyaxis}[
		axis line style = thick,
		grid = both,
		name = p1,
		% table/col sep=comma,
		width=.95\columnwidth,
		height=.6\columnwidth,
		xmin = 0, xmax=20,
		ymin = 4e-5, ymax=2,
		% ytick = {0.03,0.06,...,0.21}, yticklabels={0.03,0.06,0.09,0.12,0.15,0.18,0.21},
		% xtick = {1, 2, 3, 4}, xticklabels={0.5, 1, 1.5, 2},{font=\scriptsize},
		font=\small,
		ylabel = $\epsilon_{\sigma^2}$, xlabel = EM Iteration,
		legend style={
			font=\small,
			line width=0.3,
			nodes={scale=1.0},
		},
		legend cell align={left},
		legend pos = north east,
		]
        \addplot[color=Black, mark=*, mark options={fill=white}, very thick] 
             table[x expr=\thisrowno{0}, y expr=\thisrowno{1}] 
            {data/nse_var.txt}; \addlegendentry{Standard EM};

        \addplot[color=Cyan, mark=*, mark options={fill=white}, very thick] 
             table[x expr=\thisrowno{0}, y expr=\thisrowno{1}] 
            {data/nse_reg_var.txt};  \addlegendentry{Constrained EM};

        \addplot[name path=upper, draw=none] 
             table[x expr=\thisrowno{0}, y expr=\thisrowno{3}] 
            {data/nse_var.txt};

        \addplot[name path=lower, draw=none] 
            table[x expr=\thisrowno{0}, y expr=\thisrowno{4}] 
           {data/nse_var.txt};

        \addplot[
            fill=black!30,
            opacity=0.5
            ] fill between[of=upper and lower];

        \addplot[name path=upperReg, draw=none] 
             table[x expr=\thisrowno{0}, y expr=\thisrowno{3}] 
            {data/nse_reg_var.txt};

        \addplot[name path=lowerReg, draw=none] 
            table[x expr=\thisrowno{0}, y expr=\thisrowno{4}] 
           {data/nse_reg_var.txt}; 

        \addplot[
            fill=Cyan!30,
            opacity=0.5
            ] fill between[of=upperReg and lowerReg];

	\end{semilogyaxis}
 
% \matrix[
%     matrix of nodes,
%     anchor = north east,
%     fill = white, draw,
%     inner sep = 0.1mm,
%     column sep = 0.1mm,
%     row sep = 0.4mm,
%     node font=\footnotesize,
%     % text height=1.9ex,
%     column 1/.style={nodes={align=center}},
%     column 2/.style={nodes={align=center}},
%     column 3/.style={nodes={align=center}},
%     column 4/.style={nodes={align=center}}
%   ]
%   at ([xshift=-3pt, yshift=-3pt]current axis.north east){
%         & Perfect   & Sep. & Joi.\\ 
%     $2$dB & \ref{per2} & \ref{sep2} & \ref{joi2}\\
%     $4$dB & \ref{per4} & \ref{sep4} & \ref{joi4}\\
%     $6$dB & \ref{per6} & \ref{sep6} & \ref{joi6}\\};

\end{tikzpicture}    } 
        \vspace{-.5em}
    }
    \hspace{1em}
    \subfloat[Transition probability\label{fig:kl_ptran}]{%
        \centering
        {\begin{tikzpicture}
	
	\begin{semilogyaxis}[
		axis line style = thick,
		grid = both,
		name = p1,
		% table/col sep=comma,
		width=.95\columnwidth,
		height=.6\columnwidth,
		xmin = 0, xmax=20,
		ymin = 4e-5, ymax=2,
		% ytick = {0.03,0.06,...,0.21}, yticklabels={0.03,0.06,0.09,0.12,0.15,0.18,0.21},
		% xtick = {1, 2, 3, 4}, xticklabels={0.5, 1, 1.5, 2},{font=\scriptsize},
		font=\small,
		ylabel = $\epsilon_{P}$, xlabel = EM Iteration,
		legend style={
			font=\small,
			nodes={scale=1.0},
		},
		legend cell align={left},
		legend pos = north east,
		]

        \addplot[color=Black, mark=*, mark options={fill=white}, very thick] 
             table[x expr=\thisrowno{0}, y expr=\thisrowno{1}] 
            {data/kl_ptran.txt};  \addlegendentry{Standard EM};

        \addplot[color=Cyan, mark=*, mark options={fill=white}, very thick] 
             table[x expr=\thisrowno{0}, y expr=\thisrowno{1}] 
            {data/kl_reg_ptran.txt};\addlegendentry{Constrained EM};

        \addplot[name path=upper, draw=none] 
             table[x expr=\thisrowno{0}, y expr=\thisrowno{3}] 
            {data/kl_ptran.txt};
            
        \addplot[name path=lower, draw=none] 
            table[x expr=\thisrowno{0}, y expr=\thisrowno{4}] 
           {data/kl_ptran.txt};

        \addplot[
            fill=black!30,
            opacity=0.5
            ] fill between[of=upper and lower];

        \addplot[name path=upperReg, draw=none] 
             table[x expr=\thisrowno{0}, y expr=\thisrowno{3}] 
            {data/kl_reg_ptran.txt};

        \addplot[name path=lowerReg, draw=none] 
            table[x expr=\thisrowno{0}, y expr=\thisrowno{4}] 
           {data/kl_reg_ptran.txt};

        \addplot[
            fill=Cyan!30,
            opacity=0.5
            ] fill between[of=upperReg and lowerReg];

		\end{semilogyaxis}
 
% \matrix[
%     matrix of nodes,
%     anchor = north east,
%     fill = white, draw,
%     inner sep = 0.1mm,
%     column sep = 0.1mm,
%     row sep = 0.4mm,
%     node font=\footnotesize,
%     % text height=1.9ex,
%     column 1/.style={nodes={align=center}},
%     column 2/.style={nodes={align=center}},
%     column 3/.style={nodes={align=center}},
%     column 4/.style={nodes={align=center}}
%   ]
%   at ([xshift=-3pt, yshift=-3pt]current axis.north east){
%         & Perfect   & Sep. & Joi.\\ 
%     $2$dB & \ref{per2} & \ref{sep2} & \ref{joi2}\\
%     $4$dB & \ref{per4} & \ref{sep4} & \ref{joi4}\\
%     $6$dB & \ref{per6} & \ref{sep6} & \ref{joi6}\\};

\end{tikzpicture}    } \
        \vspace{-.5em}
    }
    \caption{Estimator performance over EM iterations for (a) NMSE of the variances and (b) KL-divergence of the transition matrix. The shaded area indicates the error region between the $25^{\text{th}}$ and $75^{\text{th}}$ percentiles. The solid circular line represents the means of the metrics across 10000 simulations. The reference and intialized IN parameters are $A=0.3$, $r=0.9$, and $\Lambda=10$ and $A=0.1$, $r=0$, and $\Lambda=1$, respectively.}
\end{figure*}
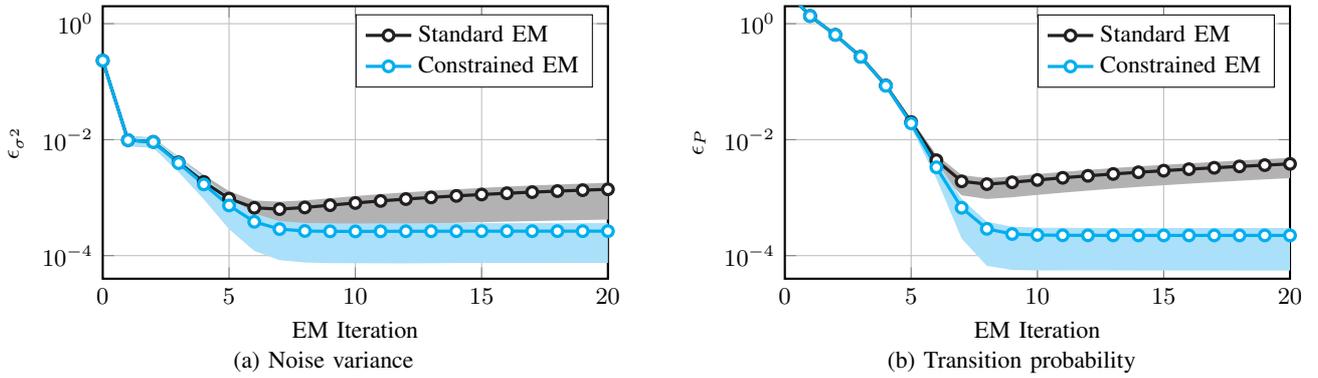

This paper aims to provide effective solutions for blind channel estimation in channels affected by bursty IN as accurate estimates of the variances and the state transition matrix are essential for MAP-based IN detectors. To address this issue, we propose a constrained EM-based algorithm. This section assesses the performance of both standard and the proposed constrained EM-based estimators adapted for BPSK transmission over a 2-state Markov-Middleton IN channel. Each transmission frame contains $32768$ bits, and we evaluate the system performance over $10000$ independent Monte Carlo simulations. For all simulations, we set the background noise power $\sigma^2_{w_0} = 1$. We assume the receivers have perfect knowledge of the Gaussian means of each state, denoted as $\mu_{s_j} \in \{-1,+1\}$. This assumption is based on the energy of BPSK symbols in the absence of ISI. Our primary goal in this paper is to assess the estimation performance regarding the variances and the state transition matrix.

\subsection{Performance metrics}
To measure the accuracy of the variance estimation, we use the normalized mean square error (NMSE) defined as 
\begin{align}
    \epsilon_{\sigma^2} =  \frac{1}{S} \sum_{j\in \mathcal{S}} \frac{(\hat{\sigma}^2_{s_j} - \sigma^2_{s_j})^2}{\sigma^4_{s_j}}. \nonumber
\end{align}
For the transition matrix, we used the Kullback–Leibler (KL) divergence to quantify the estimation performance via
\begin{align}
    \epsilon_{P} = D_{\text{KL}}(P_{s{_{ij}}} \parallel \hat{P}_{s{_{ij}}}) = \sum_{i,j} P_{s_{ij}} \log \left( \frac{P_{s_{ij}}}{\hat{P}_{s_{ij}}} \right). \nonumber
\end{align}
The number of iterations in the EM algorithm is a key factor that determines the overall computational complexity. To evaluate convergence, we examine the change in the log-likelihood of the observation sequence (model evidence), where we set a threshold
\begin{align}
    \tau = \log{p(\mathbf{y}_1^T|\theta^{(l+1)})} - \log{p(\mathbf{y}_1^T|\theta^{(l)})} = 10^{-6} \nonumber
\end{align}
to determine when the algorithm should stop iterating.

\subsection{Results}
In Figs.~\ref{fig:mse_var} and~\ref{fig:kl_ptran}, we first report the variance and transition matrix estimation performance for standard and constrained EM estimator designs regarding the number of EM iterations. The reference parameter set $\theta$ is derived from a Markov-Middleton model characterized by $A=0.3$, $\Lambda=10$, and $r=0.9$. This model is used to generate the bursty IN samples, which is shown in Fig.~\ref{fig:MMA_noise2}. To examine the situation involving incorrect assumptions about channel parameters, which often occurs in practice, we initialized our EM estimators with a relatively milder IN assumption characterized by $A=0.1$, $\Lambda=1$, and $r=0$, corresponding to the noise realizations shown in Fig.~\ref{fig:MMA_noise1}. The reference and initial assumption of the variances and transition matrix derived from the above-mentioned Markov-Middleton parameters are provided in Table~\ref{tab:para}. The results shown in Fig.~\ref{fig:mse_var} indicate that the constrained EM estimator effectively estimates the variances of the states, with performance improving over iterations and converging to an NMSE of around $3\times10^{-4}$. In contrast, although the standard EM shows significant improvement over the first six iterations, its NMSE performance does not stabilize. In the estimation of the transition matrix shown in Fig.~\ref{fig:kl_ptran}, the constrained EM estimator achieves a KL-divergence of approximately $2\times10^{-4}$, while its standard counterpart can only reach $4\times10^{-3}$, where the performance is less stable and slightly degrade after 7 EM iterations. In Table~\ref{tab:para}, we present examples of the estimates of the variances of the states and the state transition matrix for both EM estimators once the convergence condition is attained. It is important to note that while the EM estimator is guaranteed to increase the model evidence with each iteration, it does not ensure a decrease in estimation error. This has been demonstrated in the standard EM estimator, as shown in Figs.~\ref{fig:mse_var} and~\ref{fig:kl_ptran}.

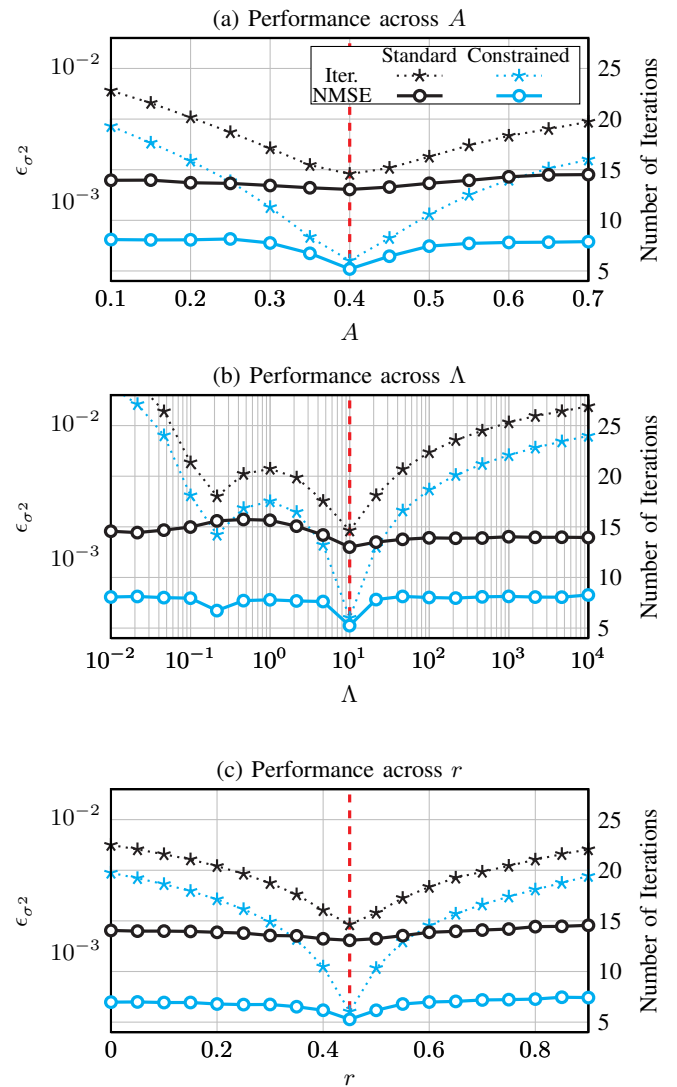
\begin{figure}[ht!]
    \centering
    \begin{subfigure}{0.49\textwidth}
        \caption{Performance across $A$}
        \begin{tikzpicture}
    \begin{axis}[ 
        axis y line*=right,
        axis line style = thick,
        % axis x line=none,
		name = p1,
        grid = both,
		% table/col sep=comma,
		width=.9\columnwidth,
		height=.55\columnwidth, 
    		xmin = 0.1, xmax=0.7,
    		ymin = 4, ymax=28,
            tick style={draw=none},
  		ytick = {5,10,15,20,25}, yticklabels={5,10,15,20,25},
            ylabel = Number of Iterations,
		font=\small,
        legend style={
            font=\footnotesize,
			nodes={scale=1.0},
		},
		legend cell align={left},
		legend pos = north east,
		]
		\addplot[color=Black, thick, dotted, mark=star, mark options={solid, fill=white, scale=1.3}] 
    		table[x expr=\thisrowno{0}, y expr=\thisrowno{5}] 
    		{data/span_A.txt}; \label{em1_it}; %\addlegendentry{IT-Stan.}; 

        \addplot[color=Cyan, thick, dotted, mark=star, mark options={solid, fill=white, scale=1.3}] 
    		table[x expr=\thisrowno{0}, y expr=\thisrowno{6}] 
    		{data/span_A.txt} ;\label{em2_it}; %\addlegendentry{IT-Cons.}; 

        \addplot[very thick, color=Red, dashed] 
            coordinates {(0.4, 0) (0.4, 30)}; 
	\end{axis}
 
	\begin{semilogyaxis}[ 
        axis line style = thick,
		width=.9\columnwidth,
		height=.55\columnwidth, 
  		xmin = 0.1, xmax=0.7,
	   	ymin = 8^-4, ymax= 8^-2,
            tick style={draw=none},
            xlabel = $A$, ylabel = $\epsilon_{\sigma^2}$,
		font=\small,
        legend style={
            font=\footnotesize,
			nodes={scale=1.0},
		},
		legend cell align={left},
		legend pos = north west,
		]
 ]
		\addplot[color=Black,  mark=*, mark options={fill=white}, very thick] 
		table[x expr=\thisrowno{0}, y expr=\thisrowno{1}] 
		{data/span_A.txt}; \label{em1_nmse}; %\addlegendentry{NMSE-Stan.}; 

        \addplot[color=Cyan,  mark=*, mark options={fill=white}, very thick] 
		table[x expr=\thisrowno{0}, y expr=\thisrowno{2}] 
		{data/span_A.txt};\label{em2_nmse}; %\addlegendentry{NMSE-Cons.} 
		
	\end{semilogyaxis}

\matrix[
    matrix of nodes,
    anchor = north east,
    fill = white, draw,
    inner sep = 0.1mm,
    column sep = 0.1mm,
    row sep = 0.5mm,
    node font=\footnotesize,
    % text height=1.9ex,
    column 1/.style={nodes={align=center}},
    column 2/.style={nodes={align=center}},
    column 3/.style={nodes={align=center}}
  ]
  at ([xshift=-3pt, yshift=-3pt]current axis.north east){
            & Standard & Constrained \\ 
    Iter.    &  \ref{em1_it}   & \ref{em2_it}  \\
    NMSE     &  \ref{em1_nmse} & \ref{em2_nmse} \\};
    
\end{tikzpicture}    

% Unified legend placemen
        \label{fig:robust_A}
    \end{subfigure}
    \begin{subfigure}{0.49\textwidth}
        \caption{Performance across $\Lambda$}
        \begin{tikzpicture}
    \begin{axis}[ 
        axis y line*=right,
        axis line style = thick,
        % axis x line=none,
		name = p1,
        grid = both,
		% table/col sep=comma,
		width=.9\columnwidth,
		height=.55\columnwidth, 
    		xmin = 10^-2, xmax=10^4,
    		ymin = 4, ymax=28,
            tick style={draw=none},
  		ytick = {5,10,15,20,25}, yticklabels={5,10,15,20,25},
            ylabel = Number of Iterations,
		font=\small,
        xmode=log,
        % xmajorgrids=true,
        ]
		\addplot[color=Black, thick, dotted, mark=star, mark options={solid, fill=white, scale=1.3}] 
		table[x expr=\thisrowno{0}, y expr=\thisrowno{5}] 
		{data/span_G.txt};

        \addplot[color=Cyan, thick, dotted, mark=star, mark options={solid, fill=white, scale=1.3}] 
		table[x expr=\thisrowno{0}, y expr=\thisrowno{6}] 
		{data/span_G.txt};
        \addplot[very thick, color=Red, dashed] 
            coordinates {(10, 0) (10, 30)}; 
	\end{axis}
 
	\begin{semilogyaxis}[ 
        axis line style = thick,
		width=.9\columnwidth,
		height=.55\columnwidth, 
  		xmin = 10^-2, xmax=10^4,
	   	ymin = 8^-4, ymax= 8^-2,
            tick style={draw=none},
            xlabel = $\Lambda$, ylabel = $\epsilon_{\sigma^2}$,
		font=\small,
        xmode=log,
 ]
		\addplot[color=Black,  mark=*, mark options={fill=white}, very thick] 
		table[x expr=\thisrowno{0}, y expr=\thisrowno{1}] 
		{data/span_G.txt};

        \addplot[color=Cyan,  mark=*, mark options={fill=white}, very thick] 
		table[x expr=\thisrowno{0}, y expr=\thisrowno{2}] 
		{data/span_G.txt};
		
	\end{semilogyaxis}

\end{tikzpicture}    
        \label{fig:robust_G}
    \end{subfigure}
    \begin{subfigure}{0.49\textwidth}
        \caption{Performance across $r$}
        \begin{tikzpicture}
    \begin{axis}[ 
        axis y line*=right,
        axis line style = thick,
        % axis x line=none,
		name = p1,
        grid = both,
		% table/col sep=comma,
		width=.9\columnwidth,
		height=.55\columnwidth, 
    		xmin = 0, xmax=0.9,
    		ymin = 4, ymax=28,
            tick style={draw=none},
  		ytick = {5,10,15,20,25}, yticklabels={5,10,15,20,25},
            ylabel = Number of Iterations,
		font=\small,
        % xmajorgrids=true,
        ]
		\addplot[color=Black, thick, dotted, mark=star, mark options={solid, fill=white, scale=1.3}] 
		table[x expr=\thisrowno{0}, y expr=\thisrowno{5}] 
		{data/span_R.txt};

        \addplot[color=Cyan, thick, dotted, mark=star, mark options={solid, fill=white, scale=1.3}] 
		table[x expr=\thisrowno{0}, y expr=\thisrowno{6}] 
		{data/span_R.txt};

        \addplot[very thick, color=Red, dashed] 
            coordinates {(0.45, 0) (0.45, 30)}; 
        
	\end{axis}
 
	\begin{semilogyaxis}[ 
        axis line style = thick,
		width=.9\columnwidth,
		height=.55\columnwidth, 
  		xmin = 0, xmax=0.9,
	   	ymin = 8^-4, ymax= 8^-2,
            tick style={draw=none},
            xlabel = $r$, ylabel = $\epsilon_{\sigma^2}$,
		font=\small,
 ]
		\addplot[color=Black,  mark=*, mark options={fill=white}, very thick] 
		table[x expr=\thisrowno{0}, y expr=\thisrowno{1}] 
		{data/span_R.txt};

        \addplot[color=Cyan,  mark=*, mark options={fill=white}, very thick] 
		table[x expr=\thisrowno{0}, y expr=\thisrowno{2}] 
		{data/span_R.txt};
		
	\end{semilogyaxis}

\end{tikzpicture}    
        \label{fig:robust_R}
    \end{subfigure}

    \caption{NMSE and the required iterations for the convergence of the variance estimation for both EM estimators across (a)~$A$, (b)~$\Lambda$, and (c)~$r$. The reference IN parameters are $A=0.4$, $r=0.45$, and $\Lambda=10$.}
\end{figure}

To investigate the robustness of the proposed EM estimators for bursty IN channels, we comprehensively evaluate the system performance across various types of CSI uncertainties in Figs.~\ref{fig:robust_A}--\ref{fig:robust_G}. In each test, we analyze the performance of the final estimates as well as the number of iterations needed for convergence. As the estimation error behavior of variances and the transition matrix are similar, we will only present the NMSE of the variance averaged over $10000$ simulations in the following analysis. We set the reference IN parameter with values $A=0.4$, $\Lambda=10$, and $r=0.45$ and initialized the system with a range of different $A$, $\Lambda$, and $r$, to represent the imperfect CSI knowledge of the probability of IN, the power of IN, and the burstiness between noise samples, respectively. Interestingly, the results shown in Figs.~\ref{fig:robust_A}--\ref{fig:robust_G} indicate that the final estimation performance is not much affected by parameter mismatches in the Markov-Middleton model. This finding implies that both EM estimators demonstrate robustness against CSI uncertainty, assuming a given IN model. However, the convergence rate significantly worsens as the level of CSI uncertainty increases, which will greatly affect the computational complexity of the system.

\section{Conclusions}\label{sec:conc}
In this paper, we tackle the challenge of channel estimation for bursty IN modeled by the Markov-Middleton channel. We developed two EM-based estimators designed explicitly for the Markov-Middleton channel model, where the variances and the transition matrix for each state are nontrivial. We show that the proposed constrained EM estimator, which exploits the knowledge of the trellis structure, can achieve a faster convergence rate and more accurate estimates compared to the standard EM approach. Our simulations show that both the standard and the constrained EM estimators are robust across a broad range of CSI uncertainties. However, the required number of iterations is greatly affected by the initialization errors. Typically, the constrained EM converges 1.5 to 2 times faster than the standard version, depending on the level of initialization errors. This advancement could pave the way for more reliable and efficient wireless communication systems in scenarios where power devices generate bursty IN, ultimately improving the overall performance of MAP-based detection techniques. In the future, it will be important to adapt the proposed EM systems for higher-order modulations and time-varying channels, as these factors are crucial for practical applications in real-world environments.

\section*{Acknowledgments}
This work was funded by the RAISE collaboration framework between Eindhoven University of Technology and NXP, including a PPS-supplement from the Dutch Ministry of Economic Affairs and Climate Policy.

\vspace{12pt}

\begin{thebibliography}{23}
\bibliographystyle{IEEEtran}

% \bibitem{Maouloud21}
% A. Maouloud, M. Klingler and P. Besnier, ``A test setup to assess the impact of EMI produced by on-board electronics on the quality of radio reception in vehicles,'' \textit{IEEE Trans. Electromagn. Compat.}, vol. 63, no. 6, pp. 1844--1855, Dec. 2021.


% \bibitem{Shan09}
% Q. Shan et al., ``Noise amplitude distribution of impulsive noise from measurements in a power substation,'' \textit{International Universities Power Engineering Conference (UPEC)}, pp. 1--5, 2009.

\bibitem{Sacuto14}
F. Sacuto, F. Labeau and B. L. Agba, ``Wide band time-correlated model for wireless communications under impulsive noise within power substation,'' \textit{IEEE Trans. Wireless Commun.}, vol. 13, no. 3, pp. 1449--1461, Mar. 2014.

\bibitem{CHC24_2}
C.-H. Chen, W.-H. Huang, B. Karanov, Y. Wu, A. Young, W. van Houtum, ``Analysis of impulsive interference in digital audio broadcasting systems in electric vehicles,'' {\it{Symposium on Information Theory and Signal Processing in the Benelux (SITB)}}, 2024. 

\bibitem{Wim22}
W. van Houtum, ``Time-division spatial interference rejection (TDSIR)-procedure,'' U.S. Patent 11,722,197B2, 2022.

\bibitem{Landa15}
I. Landa, M. M. Vélez, A. Arrinda, R. Torre and M. Fernández, ``Impulsive noise characterization and its effect on digital audio quality,'' \textit{IEEE International Symposium on Broadband Multimedia Systems and Broadcasting (BMSB)}, pp. 1--3, 2015.

\bibitem{Shan11}
Q. Shan et al., ``Estimation of impulsive noise in an electricity substation,'' \textit{IEEE Trans. Electromagn. Compat.}, vol. 53, no. 3, pp. 653--663, Aug. 2011.

\bibitem{Dario09}
D. Fertonani and G. Colavolpe, ``On reliable communications over channels impaired by bursty impulse noise,'' \textit{IEEE Trans. Commun.}, vol. 57, no. 7, pp. 2024--2030, Jul. 2009.

\bibitem{MMA}
G. Ndo, F. Labeau and M. Kassouf, ``A Markov-Middleton model for bursty impulsive noise: Modeling and receiver design,'' \textit{IEEE Trans. Power Del.}, vol. 28, no. 4, pp. 2317--2325, Oct. 2013.

\bibitem{Mitra10}
J. Mitra and L. Lampe, ``Convolutionally Coded Transmission over Markov-Gaussian Channels: Analysis and Decoding Metrics,'' \textit{IEEE Trans. Commun}, vol. 58, no. 7, pp. 1939-1949, Jul. 2010.

\bibitem{CHC24_3}
C.-H. Chen, B. Karanov, W. van Houtum, Y. Wu, and A. Alvarado, ``Turbo Receiver Design with Joint Detection and Demapping for Coded Differential BPSK in Bursty Impulsive Noise Channels,'' \textit{ArXiv preprint arXiv:2412.07911}, 2024.

\bibitem{Ndo10}
G. Ndo, P. Siohan and M. -H. Hamon, ``Adaptive Noise Mitigation in Impulsive Environment: Application to Power-Line Communications,'' \textit{IEEE Trans. Power Deliv.}, vol. 25, no. 2, pp. 647-656, Apr. 2010.

\bibitem{Oh17}
H. Oh and H. Nam, ``Design and Performance Analysis of Nonlinearity Preprocessors in an Impulsive Noise Environment,'' \textit{IEEE Trans. Veh. Technol.}, vol. 66, no. 1, pp. 364-376, Jan. 2017.

\bibitem{Alam20}
M. S. Alam, B. Selim, G. Kaddoum and B. L. Agba, ``Mitigation Techniques for Impulsive Noise With Memory Modeled by a Two State Markov-Gaussian Process,'' \textit{IEEE Syst. J.}, vol. 14, no. 3, pp. 4079-4088, Sep. 2020.

\bibitem{bcjr74}
L. R. Bahl, J. Cocke, F. Jelinek, and J. Raviv, ``Optimal decoding of linear codes for minimizing symbol error rate,'' \textit{IEEE Trans. Inform. Theory}, vol. 20, no. 2, pp. 284–287, Mar. 1974.

% \bibitem{Banelli13}
% P. Banelli, ``Bayesian Estimation of a Gaussian Source in Middleton's Class-A Impulsive Noise,'' \textit{IEEE Signal Process. Lett.}, vol. 20, no. 10, pp. 956-959, Oct. 2013.

% \bibitem{Alam18}
% M. S. Alam, G. Kaddoum and B. L. Agba, ``Bayesian MMSE Estimation of a Gaussian Source in the Presence of Bursty Impulsive Noise,'' \textit{IEEE Comm. Lett.}, vol. 22, no. 9, pp. 1846-1849, Sep. 2018.

\bibitem{Ghosh92}
M. Ghosh and C. L. Weber, ``Maximum-likelihood blind equalization,'' \textit{Optical Engineering}, vol. 31, no. 6, pp. 1224–1228, 1992.

\bibitem{Kaleh94}
G. K. Kaleh and R. Vallet, ``Joint parameter estimation and symbol detection for linear or nonlinear unknown channels,'' \textit{IEEE Trans. Commun.}, vol. 42, no. 7, pp. 2406-2413, 1994.

\bibitem{Lopes01_2}
R. R. Lopes and J. R. Barry, ``Blind iterative channel identification and equalization,'' \textit{Proc. IEEE International Conference on Communications (ICC)}, pp. 2256-2260, 2001.

% \bibitem{Selim16}
% B. Selim, O. Alhussein, S. Muhaidat, G. K. Karagiannidis, and J. Liang, ``Modeling and Analysis of Wireless Channels via the Mixture of Gaussian Distribution,'' \textit{IEEE Trans. Veh. Technol.}, vol. 65, no. 10, pp. 8309-8321, Oct. 2016

\bibitem{Schmid24}
L. Schmid, T. Raviv, N. Shlezinger, and Laurent Schmalen, ``Blind channel estimation and joint symbol detection with data-driven factor graphs,'' \textit{ArXiv preprint arXiv:2401.12627}, 2024.

\bibitem{CHC25}
C.-H. Chen, B. Karanov, I. Nikoloska, W. van Houtum, Y. Wu, and A. Alvarado, ``Modified Baum-Welch Algorithm for Joint Blind Channel Estimation and Turbo Equalization,'' \textit{International ITG Conference on Systems, Communications and Coding (SCC)}, pp. 1-6, 2025.
% C.-H. Chen, B. Karanov, I. Nikoloska, W. van Houtum, Y. Wu, and A. Alvarado, ``Modified Baum-Welch Algorithm for Joint Blind Channel Estimation and Turbo Equalization,'' \textit{ArXiv preprint arXiv:2412.07907}, 2024.

\bibitem{Zabin91}
S. M. Zabin and H. V. Poor, ``Efficient estimation of Class A noise parameters via the EM algorithm,'' \textit{IEEE Trans. Inf. Theory}, vol. 37, no. 1, pp. 60-72, Jan. 1991.

\bibitem{Awino19}
S. O. Awino, T. J. O. Afullo, M. Mosalaosi, and P. O. Akuon, ``GMM Estimation and BER of Bursty Impulsive Noise in Low-voltage PLC Networks,'' \textit{Photonics \& Electromagnetics Research Symposium - Spring (PIERS-Spring)}, pp. 1828-1834, 2019.

\bibitem{Chen19}
C.-Y. Chen and M.-C. Chiu, ``Parameter Estimation of Impulsive Noise for Channel Coded Communication Systems,'' \textit{IET Communications}, vol. 15, pp. 445-452, 2019.

\bibitem{Vannucci19}
A. Vannucci, G. Colavolpe, R. Pecori and L. Veltri, ``Estimation of a Gaussian Source with Memory in Bursty Impulsive Noise,'' \textit{IEEE International Symposium on Power Line Communications and its Applications (ISPLC)}, pp. 1-6, 2019.

\bibitem{Mirbadin21}
A. Mirbadin, A. Vannucci, G. Colavolpe, R. Pecori and L. Veltri, ``Iterative receiver design for the estimation of Gaussian samples in impulsive noise,'' Appl. Sci., vol. 11, no. 2: 557, 2021.


\bibitem{karanov24}
B. Karanov, C.-H. Chen, Y. Wu, A. Young, and W. van Houtum, ``Data-driven symbol detection for intersymbol interference channels with bursty impulsive noise,'' \textit{ArXiv preprint arXiv:2405.10814}, 2024.

\bibitem{Baum_72}
L. E. Baum, ``An inequality and associated maximization technique in statistical estimation for probabilistic functions of Markov processes,'' {\it{Inequalities III: Proceedings of the 3rd Symposium on Inequalities}}, pp. 1-8, 1972.

\bibitem{HMM}
L. R. Rabiner, ``A tutorial on hidden Markov models and selected applications in speech recognition,'' \textit{Proc. IEEE}, vol. 77, no. 2, pp. 257-286, 1989.


\bibitem{Bishop}
C. M. Bishop, \textit{Pattern Recognition and Machine Learning}, 1st ed. New York: Springer, 2006.

\end{thebibliography}
\end{document}